# A Multi Interface Grid Discovery System

A. Ali[1], A. Anjum[1, 2], J. Bunn[3], F. Khan[1], R.McClatchey[2], H. Newman[3],
C. Steenberg[3], M. Thomas[3], Ian Willers[4]

[1]*National University of Sciences and Technology, Rawalpindi, Pakistan*
[2]*CCS Research Centre, Univ. of West of England, Frenchay, Bristol BS16 1QY, UK*
[3]*California Institute of Technology, United States*
[4]*European Organization for Nuclear Research, Geneva, Switzerland*
Email:{Richard.McClatchey,Ashiq.Anjum}@cern.ch

*Abstract—* Discovery Systems (DS) can be considered as entry points for global loosely coupled distributed systems. An efficient Discovery System in essence increases the performance, reliability and decision making capability of distributed systems. With the rapid increase in scale of distributed applications, existing solutions for discovery systems are fast becoming either obsolete or incapable of handling such complexity. They are particularly ineffective when handling service lifetimes and providing up-to-date information, poor at enabling dynamic service access and they can also impose unwanted restrictions on interfaces to widely available information repositories. In this paper we present essential the design characteristics, an implementation and a performance analysis for a discovery system capable of overcoming these deficiencies in large, globally distributed environments.

## I. INTRODUCTION

With the increasingly widespread adoption of the Grid computing paradigm, there is an unprecedented increase in the number of accessible elements connected to the Grid. Services may appear or disappear: new services can become part of the Grid and older services can be withdrawn from the Grid. Moreover, the location of the services cannot be foreseen. It is therefore important to have a class of services in the distributed system that provides scalable and robust registration and discovery services. Discovery or information services enable service providers, users and applications to query for services and to retrieve up-to-date information on demand on their location and interfaces.

Most existing service discovery solutions are not suitable for the dynamic and distributed nature of the Grid, for several reasons. The main problems with current implementations are: lack of peer-to-peer discovery, lack of compliance with standards, and an absence of dynamic discovery capability for the renewal of service and multiple interfaces. Another important consideration for designing discovery services (DSs) is to build in fault tolerance mechanisms: this is especially important in a distributed system where the failure of one component can have a domino effect on others. The DS solution that we have developed is available as part of the Clarens web services framework [1], a Grid portal to run Grid-based data intensive services. Clarens is being used by different communities under the OSG (Open Science Grid) and by the NVO (National Virtual Observatory).

## II. DESIGN CHARACTERISTICS

In this section we list design requirements for building a Grid-based dynamic and distributed discovery service. The Discovery service (see figure 1) employs the Peer to Peer (P2P) paradigm in its functionality. The convergence of Grid and P2P computing offers many advantages [2], [3] such as content searches which allows P2P networks to keep a catalogue of distributed resources (files). The churn rate - the rate at which nodes leave and join the network - is much higher in such applications as compared to that of Grid. The distributed search capabilities of P2P file sharing applications can be utilized for improved performance in the Grid environment.

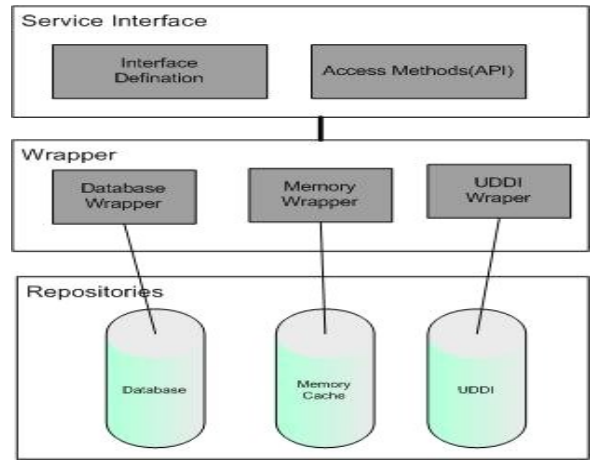

Figure 1: Discovery Service Architecture

We provide support for multiple plug-ins of different components depending upon the requirements of Grid users. The main reason for opting for a multi plug-in approach is that there are multiple applications which exhibit the same or very similar (discovery) functionality but are used by different communities. UDDI, ebXML, relational databases and LDAP can all be used as registries for storing service information and applications like MonALISA [4] or RGMA [5] etc. can be used for replication of information among the different discovery nodes. The architectural components of our discovery service can be divided into three main layers: the service interface, the wrapper and the repository. The top layer is the service interface which defines different attributes for describing a service and also methods for retrieving information from the service repository. The bottom layer (the repository) provides different variants for persistent storage of service related data. To provide an interface between different kinds of storage systems and our service interface, we provided a middle layer called the wrapper. Different sets of repositories provide more or less flexibility for storage components since each of the repositories is suitable for different types of Grid environments. Relational databases, in-memory and UDDI (Universal Description Discovery and Integration) are provided are as backend repositories.

### III. IMPLEMENTATION

A discovery service has been implemented as a web service within Clarens/JClarens. It is available in both Python (for Clarens) and Java (for JClarens [6]). Initially (Stage 1) we provided an API with the implementation of: register, deregister, find, find_key, find_server. Data on each service was stored at a centralized location. Replication was not possible between servers and consequently no communication was possible among different Clarens servers. The support for the replication of the service data was built at stage 2. We developed a new monitoring module within the MonaLisa which handles publication requess from individual instances of Clarens. Upon startup each Clarens server registered its services with the local repository and pushes them to the MonALISA server [4]. The service information is sent to any of the known station server using the ApMon client library – using simple XDR UDP packets. All distributed instances of UDDI are virtually linked to one another using the replication mechanism within our discovery service in stage 3.

### IV. PERFORMANCE ANALYSIS

We calculated the time it takes for retrieving service related data from the information service using a java based XML-RPC client. The graphs in figure 2 and 3 show the results of data retrieval with varying number of services. These results were calculated both from the memory-based and two different database-based storages. The service retrieval test does not include the time for making the connection or authenticating the user with the server. Apparently it seems that the some what larger increase in service retrieval time as number of service increase is due to the overhead involved in parsing the XML-RPC response from the Clarens server. We also noted that the in-memory cache is suitable for fast retrieval of services data. The only problem is with the memory over flow when we register a large number of services or data. Another test was to measure the time the information service takes to replicate service information to other instances of information services over the wide area network. The Service retrieval time was calculated based on the difference in time between a service being registered at one node of JClarens and becoming available at another node. Figure 4 presents the values obtained for service retrieval for different number of attempts. The variance in the upper values is attributed to the network latency prevalent in our network.

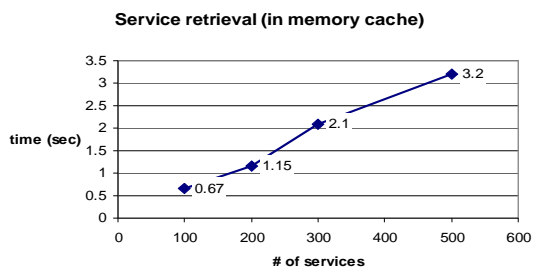

Figure 2: Service retrieval (in memory cache)

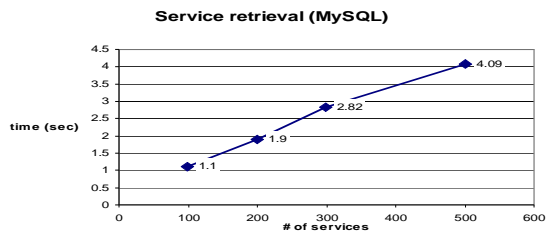

Figure 3: Service retrieval (MySQL)

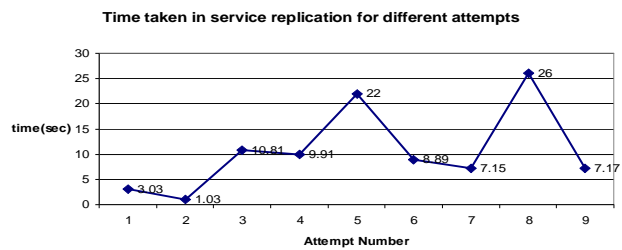

Figure 4: Service replication

### V. CONCLUSIONS

This paper describes the design, implementation and performance analysis of the discovery and information system. We presented how our system tackles the most demanding issues of service registration and life time, information propagation, replication and service discovery The ability to attach arbitrary set of key-value pairs with service information makes it possible to easily extend/customize the service description interface. An analysis is presented to to gauge how our system behaves with changing numbers of services.

### REFERENCES

[1] F. van Lingen et al. "The Clarens Web Service Framework for Distributed Scientific Analysis in Grid Projects", ICPP 2005
[2] I Foster, A. Iamnitchi, "On Death, Taxes, and the Convergence of Peer-to-Peer and Grid Computing". Lecture Notes in Computer Science, 2003. Springer Verlag
[3] Karan Bhatia. "Peer-To-Peer Requirements On The Open Grid Services Architecture Framework". GFD-I.049 OGSAP2P Research Group.
[4] I. Legrand et al. "MonALISA: an Agent Based, Dynamic Service System to Monitor, Control and Optimize Grid Based Applications", CHEP, Interlaken, Switzerland, 2004.
[5] AW Cooke et al.,"The Relational Grid Monitoring Architecture: Mediating Information about the Grid"- Journal of Grid Computing (2004) 2: 323–339 © Springer 2005
[6] M. Thomas et al., "JClarens: A Java Framework for Developing and Deploying Web Services for Grid Computing", ICWS 2005, Florida, 2005, ISBN 0-7695-2409-5 pp141-148